\documentclass[a4paper]{jpconf}

\usepackage{cite}

\usepackage{multirow}
\usepackage{graphicx}
\usepackage[abs]{overpic}
\usepackage{dcolumn}
\usepackage{bm}
\usepackage{color}

\begin{document}

\title{Coulomb excitation of exotic nuclei at the R$^3$B-LAND setup}


\author{D~M~Rossi$^{1,2}$, P~Adrich$^1$, F~Aksouh$^{1,}$\footnote{Present address: King Saud University, Riyadh, Saudi Arabia.}, H~Alvarez-Pol$^3$, T~Aumann$^{1,4}$, J~Benlliure$^3$, M~B\"ohmer$^5$, K~Boretzky$^1$, E~Casarejos$^6$, M~Chartier$^7$, A~Chatillon$^1$, D~Cortina-Gil$^3$, U~Datta~Pramanik$^8$, H~Emling$^1$, O~Ershova$^9$, B~Fernandez-Dominguez$^{3,7}$, H~Geissel$^1$, M~Gorska$^1$, M~Heil$^1$, H~Johansson$^{1,10}$, A~Junghans$^{11}$, O~Kiselev$^{1,2}$, A~Klimkiewicz$^{1,12}$, J~V~Kratz$^2$, N~Kurz$^1$, M~Labiche$^{13}$, T~Le~Bleis$^{1,9,14}$, R~Lemmon$^{15}$, Yu~A~Litvinov$^1$, K~Mahata$^{1,16}$, P~Maierbeck$^5$, A~Movsesyan$^4$, T~Nilsson$^{10}$, C~Nociforo$^1$, R~Palit$^{17}$, S~Paschalis$^{4,7}$, R~Plag$^{1,9}$, R~Reifarth$^{1,9}$, H~Simon$^1$, K~S\"ummerer$^1$, A~Wagner$^{11}$, W~Walus$^{12}$, H~Weick$^1$ and M~Winkler$^1$}

\address{$^{1}$ GSI Helmholtzzentrum f\"ur Schwerionenforschung GmbH, Darmstadt, Germany}
\address{$^{2}$ Johannes Gutenberg-Universit\"at, Mainz, Germany}
\address{$^{3}$ University of Santiago de Compostela, Santiago de Compostela, Spain}
\address{$^{4}$ Technische Universit\"at Darmstadt, Darmstadt, Germany}
\address{$^{5}$ Technische Universit\"at M\"unchen, Garching, Germany}
\address{$^{6}$ University of Vigo, Vigo, Spain}
\address{$^{7}$ University of Liverpool, Liverpool, United Kingdom}
\address{$^{8}$ SINP, Kolkata, India}
\address{$^{9}$ Goethe Universit\"at, Frankfurt am Main, Germany}
\address{$^{10}$ Chalmers University of Technology, G\"oteborg, Sweden}
\address{$^{11}$ Helmholtz-Zentrum Dresden-Rossendorf e.V., Dresden, Germany}
\address{$^{12}$ Jagiellonian University, Krakow, Poland}
\address{$^{13}$ University of the West of Scotland, Paisley, United Kingdom}
\address{$^{14}$ IPHC, Strasbourg, France}
\address{$^{15}$ STFC Daresbury Laboratory, Daresbury, United Kingdom}
\address{$^{16}$ BARC, Mumbai, India}
\address{$^{17}$ TIFR, Mumbai, India}

\ead{d.rossi@gsi.de}

\begin{abstract}
Exotic Ni isotopes have been measured at the R$^{3}$B-LAND setup at GSI in Darmstadt, using Coulomb excitation in inverse kinematics at beam energies around 500~MeV/u. As the experimental setup allows kinematically complete measurements, the excitation energy was reconstructed using the invariant mass method. The GDR and additional low-lying strength have been observed in $^{68}$Ni, the latter exhausting 4.1(1.9)\% of the E1 energy-weighted sum rule. Also, the branching ratio for the non-statistical decay of the excited $^{68}$Ni nuclei was measured and amounts to 24(4)\%.
\end{abstract}

\section{Introduction}

Unstable nuclei have been shown to exhibit several exotic features that cannot be observed in their stable counterparts. In neutron-rich nuclei, one of these features is the appearance of electric-dipole strength at energies near the neutron separation threshold. Located below the well-known Giant Dipole Resonance (GDR), which has already been studied extensively in stable species~\cite{die1988}, this new low-lying E1 strength distribution is often denoted as Pygmy Dipole Resonance (PDR), referring to its much smaller photoabsorption strength compared to that of the GDR. Despite its small E1 strength, the PDR provides an experimental access to the equation-of-state (EOS) of asymmetric nuclear matter, and thus a link to the neutron skin evolution~\cite{pie2011}, as well as about the abundance distribution of heavy elements in the universe. Indeed, these two phenomena arise from the behavior of the EOS of nuclear matter, more precisely of its symmetry-energy term. By constraining the various parameters of the EOS using measurements of the PDR, not only the global description of nuclear matter under various conditions (asymmetry and density) can be improved, but also the size, structure and chemical composition of neutron stars can be evaluated in a more realistic manner. The description of the abundance of heavy nuclei, on the other hand, can be directly influenced by the spectral shape of the low-lying E1 strength~\cite{gor1998}, affecting the particle (neutron or proton) capture and photodissociation reactions in various nucleosynthesis scenarios.

One of the first PDR observations was realized at the LAND setup by studying the Coulomb excitation $\left( \gamma^{*},xn\right)$ of neutron-rich Sn isotopes~\cite{adr2005}. The PDR was measured in $^{130}$Sn and $^{132}$Sn at energies around 10~MeV and exhausted 7(3)\% and 4(3)\% of the E1 energy-weighted sum rule, respectively. The comparison of these results with RPA calculations~\cite{kli2007} served to constrain the symmetry energy and the symmetry-energy pressure, as well as the neutron-skin thicknesses of the studied Sn isotopes and of $^{208}$Pb. More recently, the low-lying E1 strength in $^{68}$Ni was measured with $\left( \gamma^{*},\gamma'\right) $ virtual photon scattering~\cite{wie2009}, revealing 5\% of the E1 energy-weighted sum-rule strength under the assumption of a branching ratio for the direct $\gamma$-decay of $\sim 4$\% in the PDR region. From these results, Carbone \textit{et al.}~\cite{car2010} derived symmetry-energy parameters that are in good agreement with those previously obtained from $^{130,132}$Sn.

The scarce experimental data on low-lying E1 strength in exotic nuclei leaves many important questions related to the PDR open. For instance, is there a systematic dependence of the PDR energy, width or strength on variables such as the mass, charge or isospin asymmetry? Also, does the decay of the PDR significantly differ from that of the GDR, giving, \textit{e.g.}, enhanced direct $\gamma$ or direct neutron decay branching ratios? What is the degree of collectivity of the PDR, and is it's character rather isovector or isoscalar? All these questions require systematic studies of the low-lying E1 strength to be carried out over various mass and isospin-asymmetry ranges, in which quantities, such as the differential photoabsorption cross section or the dipole polarizability, are measured and can then be compared to theory.

We present here an experiment in which the E1 strength has been measured in neutron-rich Ni isotopes by means of electromagnetic excitation in inverse and complete kinematics. After the description of the R$^{3}$B-LAND experimental setup and of the measurement principle, preliminary results for $^{68}$Ni will be presented.

\section{Experimental setup}\label{sec:setup}

The R$^{3}$B-LAND setup is located at GSI Helmholtzzentrum f\"ur Schwerionenforschung GmbH in Darmstadt, Germany, and allows the investigation of reactions in inverse kinematics involving very short-lived nuclei. Stable primary ion beams are prepared and accelerated to relativistic energies in the UNILAC linear accelerator combined with the SIS-18 heavy-ion synchrotron. The production of unstable secondary beams takes place in the fragment separator FRS~\cite{gei1992}, consisting of a production target and of a series of dipole magnets allowing for the in-flight separation of the produced species. In the present case of the investigation of $^{68}$Ni, a stable $^{86}$Kr primary beam was accelerated to approximately 650~MeV/u and directed onto a 4.2~g/cm$^{2}$ Be target, generating multiple secondary species by projectile fragmentation at the entrance of the FRS. Within two distinct settings of the FRS, the Ni isotopic chain was covered from A=67 to A=72, with fragment energies of the order of 500~MeV/u. The exotic secondary beams contained a mixture of several nuclei of interest delivered simultaneously, therefore requiring an eventwise particle identification, making use of three measurements at the FRS: a precise measurement of the $B\rho$ value for each particle, its velocity from a time-of-flight measurement and its charge, obtained from the energy loss.

The R$^{3}$B-LAND setup was arranged in the present case to allow a kinematically complete measurement of electromagnetic excitation reactions. Due to the high beam velocities of the order of $\beta \approx 0.75$, all the reaction products are strongly forward-focused. In most cases, this effect is beneficial, since almost full solid-angle coverage can be achieved with reasonably-sized detectors. In other cases, however, the kinematical boost induces an additional difficulty for the detection and data analysis, such as for the Doppler shift in the photon decay of excited reaction fragments.
\begin{figure}
 \centering
 \includegraphics[width=\textwidth,bb=0 0 800 600]{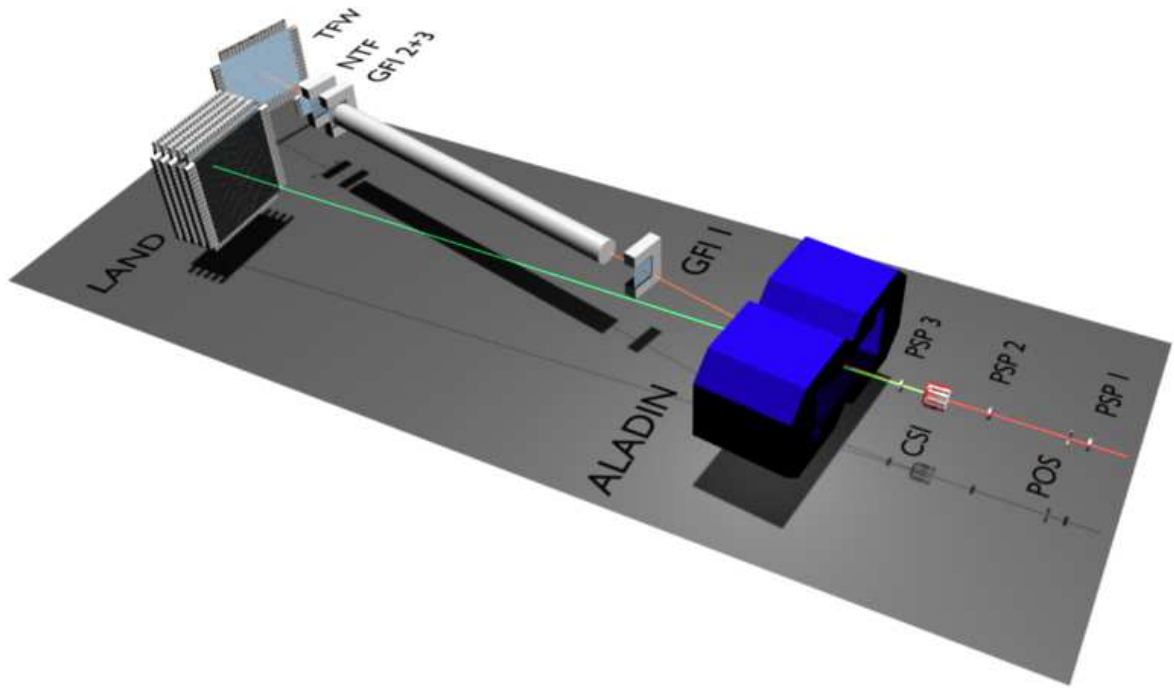}
 \caption{(Color online) R$^{3}$B-LAND setup for the Ni experiment at GSI, Darmstadt. See text for description.}
 \label{fig:land_setup}
\end{figure}
Figure~\ref{fig:land_setup} shows the R$^{3}$B-LAND setup used for the present experiment on neutron-rich Ni isotopes. The secondary beam enters the setup from the lower right side of the figure, and passes through a series of tracking and identification detectors. Two silicon pin-diodes (PSP1/2) provide the energy-loss and position measurements, while the plastic scintillator detector POS generates a signal used as minimum-bias trigger and as timing reference for all other detectors. The reaction target is located directly in front of the CsI gamma detector, a $2\pi$ barrel-type detector consisting of 144 crystals arranged in 12 azimuthal and 12 polar segments, detecting photons emitted by the excited beam fragments. The fine segmentation was designed to provide a relatively small Doppler broadening, thus increasing the energy resolution of the detector. Heavy charged particles, both non-reacted incoming species as well as reaction fragments, are strongly forward-focused and are detected by a third pin-diode (PSP3), providing a first energy-loss measurement of the reaction products and a position measurement. These charged particles are then deflected in the ALADIN dipole magnet into a fragment branch equipped with three plastic scintillating fiber detectors (GFI1-3)~\cite{cub1998} and a plastic scintillator time-of-flight wall (TFW). These detectors provide position, time-of-flight and energy-loss measurements, from which the fragment mass can be reconstructed. Neutrons evaporated during the decay process of the excited nuclei also have very small laboratory angles, allowing them to reach the Large Area Neutron Detector LAND~\cite{bla1992}, located at $0^{\circ}$ with respect to the incoming beam axis and approximately 15~m downstream of the target.

Coulomb excitation is performed preferentially using a high-Z target, such as $^{nat}$Pb in the present experiment.
\begin{figure}
 \centering
 \includegraphics[width=0.7\textwidth,bb=0 0 289 182]{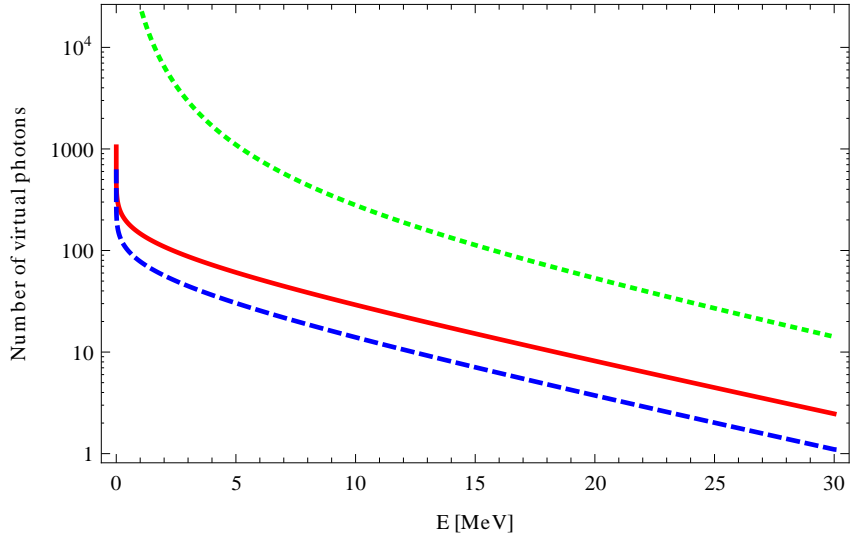}
 \caption{(Color online) Virtual photon spectra for $^{68}$Ni at 500~MeV/u on a $^{nat}$Pb target. Virtual photon numbers for E1 (solid red line), E2 (dotted green line) and M1 (dashed blue line) photons.}
 \label{fig:ni68_vps}
\end{figure}
Figure~\ref{fig:ni68_vps} displays the number of virtual photons for as a function of their energy~\cite{ber1986}, representing the availability of a given excitation energy. This particular case was calculated for a $^{68}$Ni beam impinging on a $^{nat}$Pb target at 500~MeV/u, with a minimum impact parameter of 12.98~fm, obtained with the parameterization of Benesh, Cook and Vary~\cite{ben1989} based on the nuclear radii of the collision partners. The nature of the Coulomb excitation process prevents us from knowing the excitation energy \textit{a priori}. Therefore, it can only be reconstructed using the invariant-mass technique, in which the four-momenta of all species participating in the reaction must be measured. The reconstructed excitation energy can be expressed as:
\begin{equation}\label{eq:estar}
 E^{*}=\sqrt{\sum_{i}m_{i}^{2}+\sum_{i\neq j}\gamma_{i}\gamma_{j}m_{i}m_{j}\left( 1-\beta_{i}\beta_{j}\cos\vartheta_{ij}\right) } + E_{\gamma,sum} - m_{proj}
\end{equation}
where the indices $i$ and $j$ both run over all outgoing species and $E_{\gamma,sum}$ is the total energy carried by photons. It should be noted that equation~\ref{eq:estar} is only valid in this form when all quantities are expressed in the incoming projectile's rest frame. While choosing an appropriate target nucleus to maximize the probability of electromagnetic excitation, a non-negligible probability for nuclear excitation remains, as well as a reaction-like signature induced by background interactions of the beam particles. As the differential cross section measured with the Pb target can be approximated as a linear combination of Coulomb, nuclear and background contributions, each nucleus of interest must be measured on three different targets, namely $^{nat}$Pb, $^{nat}$C and an empty target. These three measurements allow a separation of the various contributions, providing clean Coulomb excitation differential cross sections.

\section{Evaluation of experimental response}\label{sec:resp}

The comparison of theoretical differential cross sections with the experimental distributions requires an accurate understanding of the response function induced by the detectors of the setup and by the algorithms used to treat the measured data. Mainly two detectors of the setup are responsible for the response, namely LAND and the CsI detector. The neutron reconstruction algorithm of LAND processes the observed hits of each event into neutron hits using a series of conditions on the time and position variables. Due to this algorithm, there is a non-negligible probability of not detecting the proper number of neutrons. Additionally, the energy-dependent efficiency and acceptance must also be taken into account at this level. In the case of the CsI detector, the main response contribution lies in the limited calorimetric properties of the detector, thus shifting the real photon energies to much lower values. All of the effects cited above are evaluated quantitatively using a proper decay simulation for the nucleus of interest linked to a response simulation, taking into account all detector-related effects. While the CsI response is mainly simulated using GEANT4~\cite{ago2003} within the R3BROOT framework~\cite{ber2008}, LAND relies on a different simulation principle. Instead of simulating completely the response of the neutron events with a simulation package (\textit{e.g.}, GEANT4), the neutrons from the event generator are projected onto the current LAND configuration, taking its distance to the target and eventual inactive sub-modules into account. For each neutron hitting LAND, a previously measured neutron hit distribution, including all secondary particles, is retrieved from a database and shifted to the simulated interaction point. All neutron hits for a given event are processed in this manner, before the complete hit distribution is evaluated by the neutron reconstruction algorithm. The neutron hit distributions in the database have been previously measured with LAND and consist of real 1-neutron events from a deuteron break-up experiment. This procedure allows us to be independent of any simulation package for the description of the interaction of high-energy neutrons with matter.

Based on this simulation of the detector response, we are able to compare any theoretical input with measured data. This requires the convolution of the theoretical input according to the following expression for, \textit{e.g.}, the excitation energy:
\begin{equation}\label{eq:conv}
 \left( \frac{d\sigma}{dE}\right)_{r} \left( E_{r}\right) = \sum_{i=1}^{4}\sum_{j=1}^{3}\int_{E_{g}} \mathcal{R}\left( i,j,E_{g},E_{r}\right)\cdot \mathcal{B}\left( i,E_{g}\right)\cdot \mathcal{M}\left( i,j\right) \cdot \left( \frac{d\sigma}{dE}\right)_{g} \left( E_{g}\right) dE_{g}
\end{equation} 
where $E_{g}$ is the generated excitation energy, $E_{r}$ is the reconstructed excitation energy, $i$ and $j$ are the number of generated and detected neutrons, respectively, $\mathcal{R}$ is the response matrix, $\mathcal{B}$ is the branching ratio matrix and $\mathcal{M}$ is the mass separation matrix. The observable $E$ of Eq.~\ref{eq:conv} can be exchanged freely for any other observable of interest. While the matrices $\mathcal{R}$ and $\mathcal{B}$ are obtained from the statistical decay simulation, the mass separation matrix $\mathcal{M}$ is defined by the experimental conditions. This matrix takes into account the eventual limited mass separation capability of the setup, therefore mixing data from reaction channels involving neighboring fragment masses with the channel of interest. In the present case, the mass resolution obtained in the experiment is sufficient to exclude contaminants by mass cuts, such that the mass separation matrix becomes:
\begin{equation}\label{eq:mass_matr}
        \mathcal{M}\left(i,j\right) = \left(
\begin{array}{ccc}
        1 & 0 & \ldots\\
        0 & 1 & \ldots\\
        \vdots &\vdots &\ddots
\end{array}
\right)
\end{equation}
The matrix $\mathcal{B}\left( i,E_{g}\right)$ consists of the branching ratio for channel $^{A}$X($\gamma^{*}$,$i$n)$^{A-i}$Y as a function of the generated excitation energy $E_{g}$ (or any other observable used in Eq.~\ref{eq:conv}).

The description of the response matrix $\mathcal{R}\left( i,j,E_{g},E_{r}\right)$ is somewhat more complicated: it relates two observables, one generated and one reconstructed, for a given generated channel $^{A}$X($\gamma^{*}$,$i$n)$^{A-i}$Y and for a given number of reconstructed neutrons $j$. Since in the present case the mass separation matrix does not contain off-diagonal elements, we only need response matrices where $i=j$, since all other combinations of $i$ and $j$ are rejected.
\begin{figure}[tb]
 \centering
 \includegraphics[width=\textwidth,bb= 0 0 567 236]{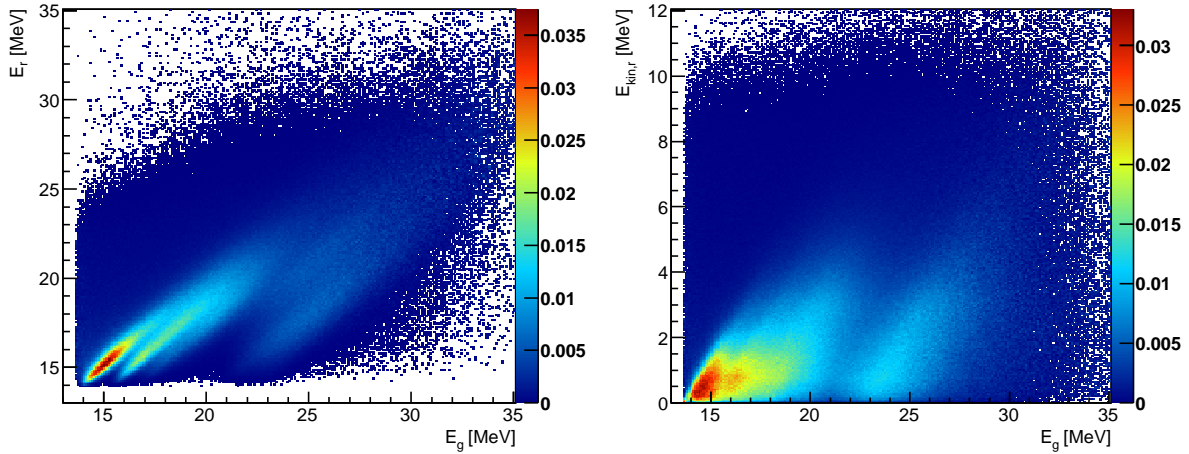}
 \caption{(Color online) Response matrices for the $^{68}$Ni($\gamma^{*}$,2n)$^{66}$Ni channel, relating the reconstructed excitation energy $E_{r}$ (left) and reconstructed neutron kinetic energy $E_{kin,r}$ (right) with the generated excitation energy $E_{g}$.}
 \label{fig:rmat}
\end{figure}
Figure~\ref{fig:rmat} shows two response matrices for the $^{68}$Ni($\gamma^{*}$,2n)$^{66}$Ni channel ($i=j=2$) for the reconstructed excitation energy $E_{r}$ (left) and reconstructed neutron kinetic energy $E_{kin,r}$ (right) as a function of the generated excitation energy $E_{g}$. The response matrix for the excitation energy nicely illustrates the strong distortion induced by the CsI detector. An ideal setup, consisting of fully efficient detectors and without any distortions due to the response, one would expect a strong linear correlation lying on the diagonal $E_{r}=E_{g}$ line of the response matrix. This is indeed the case for small excitation energies (from S$_{2n}$ up to $\sim$15.5 MeV). At larger energies, however, the correlation on the diagonal line disappears in favor of a second correlation band with an offset to lower $E_{r}$ values. At even higher energies, more such correlation bands appear. This is due to the limited calorimetric efficiency of the CsI detector, as soon as photon emission is involved in the decay of the excited nucleus. In the case of $\gamma$-cascades (\textit{e.g.}, above 22~MeV in the left part of Fig.~\ref{fig:rmat}), at least two secondary bands are visible, each one representing combinations of missed individual photons.

Due to the complicated structure of the response matrices, a deconvolution of the measured data with the inverted response is not feasible. In order to compare a theoretical input with the experimental data, an iterative scheme based on the convolution of the theoretical input must be used. At the beginning of each iteration, a trial input, usually representing the shape of the expected E1 strength, is convoluted according to Eq.~\ref{eq:conv}, producing the reconstructed differential cross section based on the input distribution. In the second step, the convoluted trial function is compared to the experimental data by calculating the $\chi^{2}$ value. The last step consists of an adjustment of the trial input in order to minimize the $\chi^{2}$ value. The iterative procedure continues until a $\chi^{2}$ minimum is reached, at which point one considers that the trial input provides a possible deconvoluted description of the experimental data.

\section{Neutron decay of $^{68}$Ni}\label{sec:n_decay}

In the present experiment, the measured E1 strength was limited to excitation energies above the neutron threshold of the nucleus of interest. Therefore, several decay channels are open, such as the direct photon, direct particle and compound nucleus decays. When studying giant resonances in heavy nuclei, it is usually safe to assume that the compound nucleus decay is the dominant channel. Statistical models are capable of describing quantitatively the kinetic-energy distributions of particles emitted by an evaporation-like decay of the compound nucleus. In the case of one-neutron emission from a nucleus A, these distributions follow Maxwell distributions~\cite{wei1937}:
\begin{equation}\label{eq:maxwell}
 \frac{d\sigma}{dE_{kin}} \left( E_{kin}\right) \propto \sigma_{c}\left( E_{kin}\right) E_{kin}\exp \left[ -\frac{E_{kin}}{T}\right] 
\end{equation} 
where $\sigma_{c}$ is the neutron capture cross section on nucleus A-1 and $T$ is the nuclear temperature of the daughter nucleus. Since Eq.~\ref{eq:maxwell} is only valid for decays involving one neutron, Le Couteur and Lang~\cite{lec1959} provide an expression valid for $n>1$ emitted neutrons:
\begin{equation}\label{eq:maxwell_mod}
 \frac{d\sigma}{dE_{kin}} \left( E_{kin}\right) \propto \sigma_{c}\left( E_{kin}\right)\frac{E_{kin}^{l-1}}{T_{*}^{l}} \exp \left[ -\frac{E_{kin}}{T_{*}}\right] 
\end{equation} 
where $l\approx\frac{16}{11}$ and $T_{*}\approx\frac{11}{12}T$. The substantial difference between expressions~\ref{eq:maxwell} and \ref{eq:maxwell_mod} resides in the pre-factor of the exponential term. One therefore expects a pre-factor $\sim E_{kin}$ in the 1n decay channel, and $\sim\sqrt{E_{kin}}$ for all other neutron decay channels.

The quantitative understanding of the experimental neutron kinetic-energy distributions requires a simulation of the decay of the compound nucleus coupled to a simulation of the detector response, as described in section~\ref{sec:resp}. The generation of neutron energies in this process is, however, difficult, since for a given excitation energy, the number of emitted neutrons is often not known \textit{a priori}.
\begin{figure}[tb]
 \centering
 \includegraphics[width=0.4\textwidth,bb=0 0 732 891]{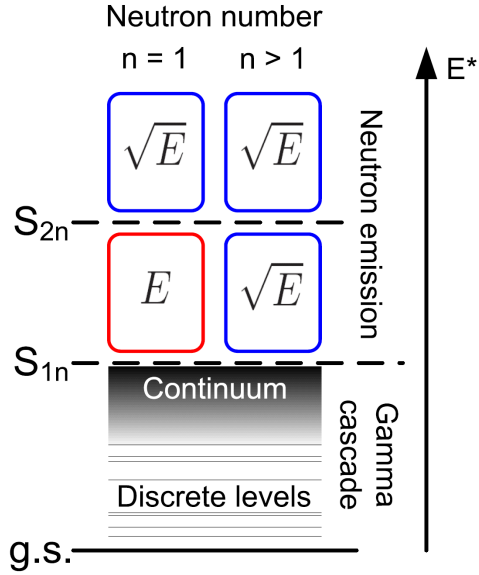}\hspace{.2 cm}
 \begin{minipage}[b]{0.4\textwidth}
  \caption{\label{fig:ekin_algo}(Color online) Visualization of the algorithm used to generate neutron kinetic energies. The pre-factor of the Maxwellian distribution is shown for the $n^{th}$ neutron being processed by the event generator.}
 \end{minipage} 
\end{figure}
Figure~\ref{fig:ekin_algo} shows graphically how the algorithm chooses the Maxwell pre-factor. Only if the first candidate for the neutron evaporation is being processed and only if the excitation energy of the parent nucleus is located between the 1n and 2n thresholds the $E_{kin}$ pre-factor is chosen. In all other cases, $\sqrt{E_{kin}}$ is used. In order to calculate the kinetic-energy distributions, additional information concerning $T$ is required. The temperature of a nucleus is usually defined as $T=\sqrt{U/a}$, where $U$ is the internal excitation energy and $a$ is the level-density parameter of the nucleus. Evaluating $a$ is not straightforward, however, several parameterizations exist.
\begin{figure}
 \centering
\begin{minipage}[t]{0.48\textwidth}
\includegraphics[width=\textwidth,bb= 0 0 567 514]{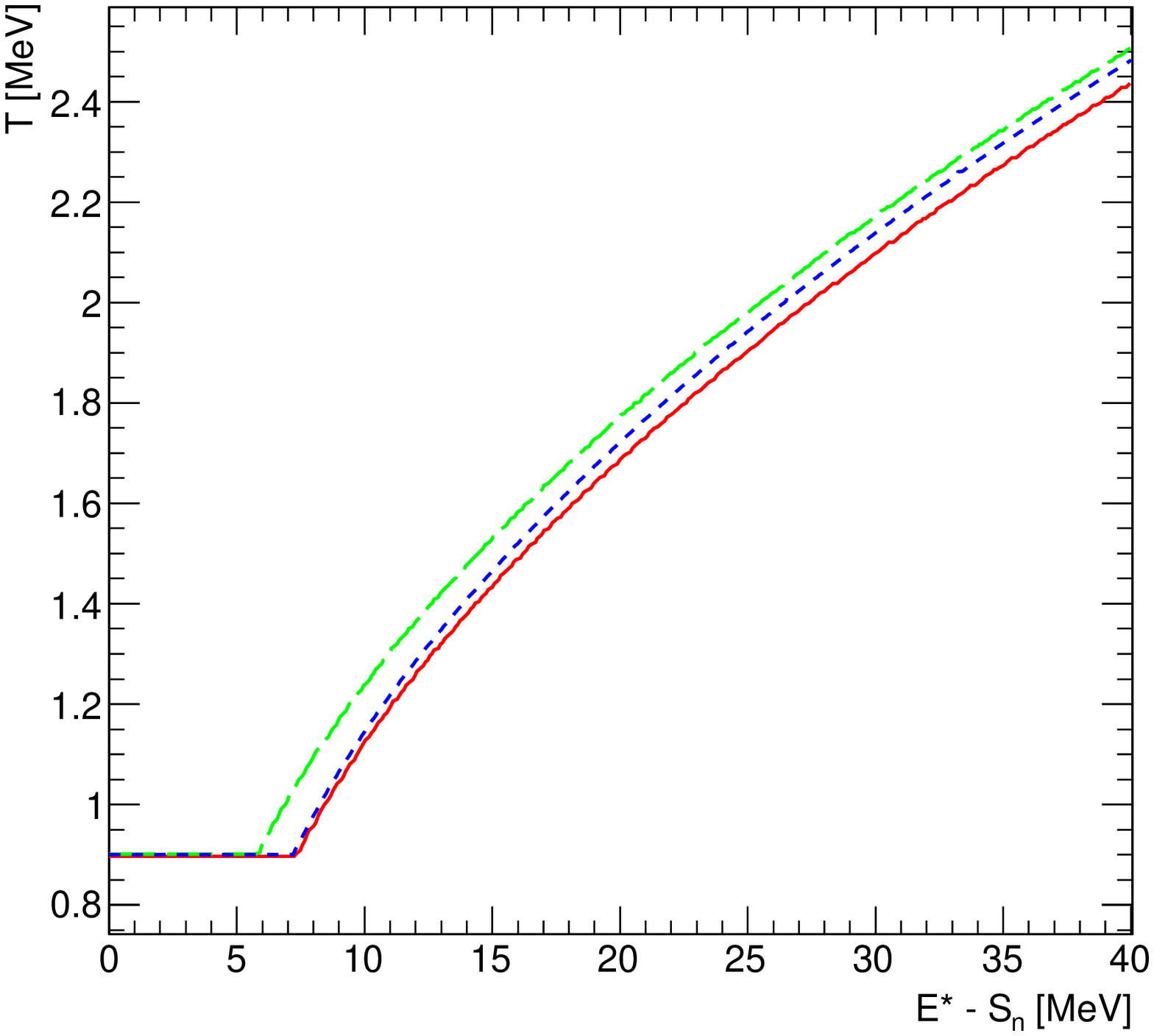}
  \caption{(Color online) Temperature profiles for $^{68}$Ni (solid red line), $^{67}$Ni (dashed green line) and $^{66}$Ni (dotted blue line) nuclei.}
 \label{fig:tempni}
\end{minipage} 
\hspace{.2 cm}
\begin{minipage}[t]{0.48\textwidth}
  \begin{overpic}[scale=0.351]{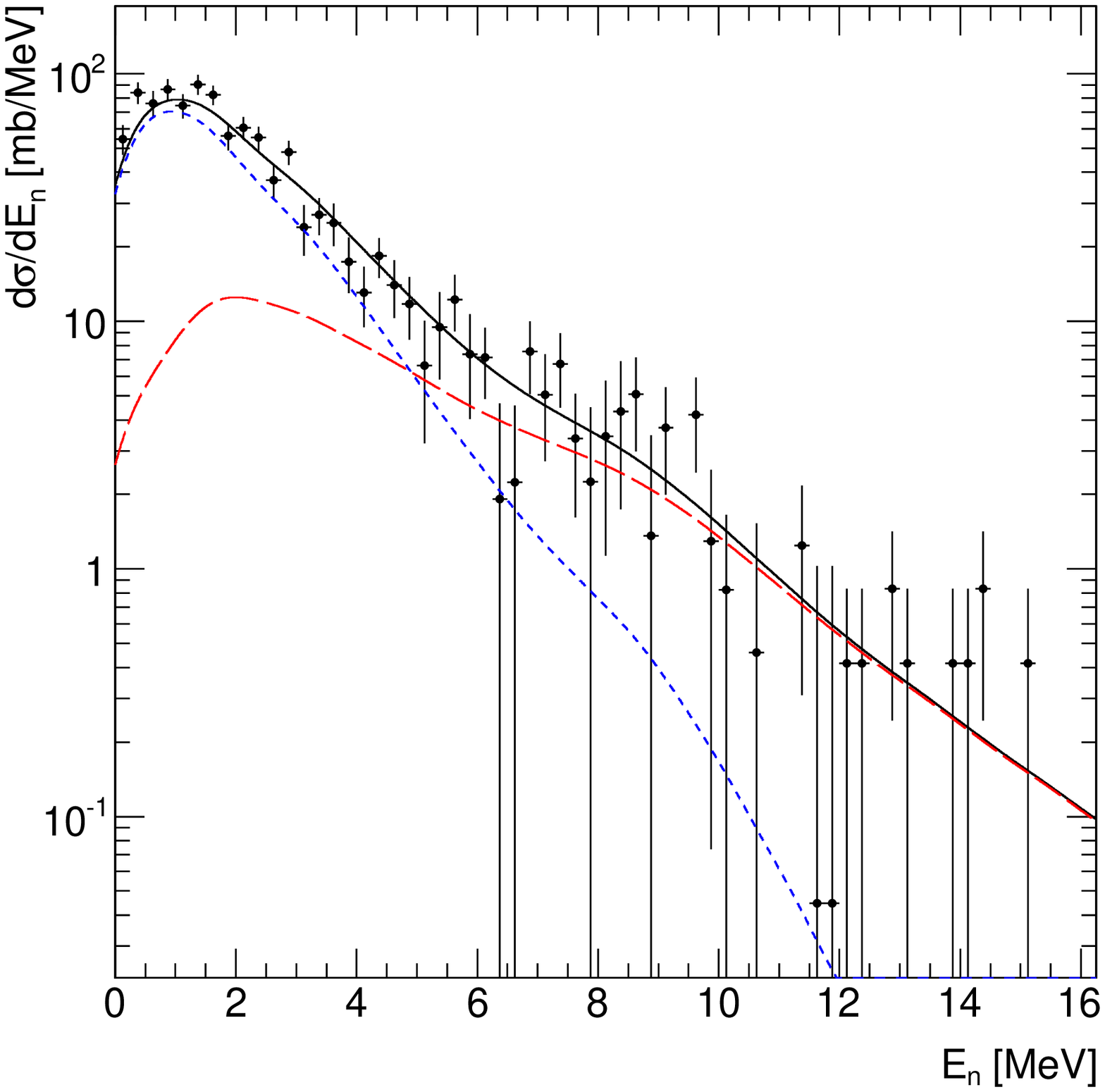}
  \put(40,55){Preliminary}
  \end{overpic} 
  \caption{(Color online) Neutron kinetic energy-differential cross section for the $^{68}$Ni($\gamma^{*}$,n)$^{67}$Ni channel. The experimental data (black points), the total neutron kinetic-energy fit (solid black line), the statistical decay component (dotted blue line) and the non-statistical decay component (dashed red line) are shown.}\label{fig:ekin} 
\end{minipage} 
\end{figure}
In the present case, we chose to use ABRABLA~\cite{kel2008,gai1991} to evaluate the nuclear temperature profiles for various nuclei of interest. Figure~\ref{fig:tempni} presents the temperature of $^{66-68}$Ni as a function of excitation energy calculated in this manner. The curves are directly used by the statistical decay algorithm, allowing it to directly generate neutron energies according to Eq.~\ref{eq:maxwell} and \ref{eq:maxwell_mod}.

One can now attempt to describe quantitatively the energy-differential cross section of the evaporated neutrons in the $^{68}$Ni($\gamma^{*}$,n)$^{67}$Ni channel, as shown in Fig.~\ref{fig:ekin}. The black points represent the measured data, obtained from the time-of-flight measurement between the target and LAND, as well as the position of the reconstructed neutron in the latter detector. As described in section~\ref{sec:setup}, the differential Coulomb-excitation cross section is obtained by measuring the reaction on a $^{nat}$Pb, a $^{nat}$C and an empty target in order to disentangle the electromagnetic, nuclear and background components. While the background subtraction is straightforward and is performed on the reaction-probability level, the subtraction of the nuclear contribution requires the rescaling of the cross section measured on the carbon target for direct use on the lead data, using a semi-empirical scaling function based on the radii of the target nuclei.

Before considering the curves obtained from the fitting procedure, one clearly sees that the experimental distribution extends significantly beyond neutron energies at which the 2n decay channel is open. Since for $^{68}$Ni the 1n and 2n thresholds are located at 7.792~MeV and 13.60~MeV, respectively, the 2n channel opens at 5.81~MeV above the 1n threshold. It is therefore not surprising that the data above 5.81~MeV cannot be described by the statistical decay simulation alone, shown as the dotted blue line in Fig.~\ref{fig:ekin}. Since a different decay mode is required to describe the high-energy portion of the kinetic energies, the direct neutron decay will be investigated. As a rough assumption, only the direct decay into the ground state of the A-1 nucleus is considered. The decay is easily simulated, since the neutron carries away the entire excitation energy minus its binding energy. The simulation of the detector response remains the same as in the case of the statistical decay. The obtained simulation data is shown in Fig.~\ref{fig:ekin} as the dashed red line, mainly describing the experimental data above 7~MeV. The sum of both contributions is represented by the solid black line, describing well the entire differential cross section distribution. A $\chi^{2}$ minimization of the sum of the statistical and direct decay contributions leads to a direct decay branching ratio of 24(4)\%, which is in good agreement with values obtained for other nuclei in this mass region~\cite{har2001}.

\section{E1 strength in $^{68}$Ni}

The excitation energy-differential cross section of $^{68}$Ni is reconstructed using Eq.~\ref{eq:estar} for the 1n and 2n data. Since the 3n threshold is located at 22.55~MeV, the analysis of the 1n and 2n channels covers most of the energy range of the GDR and of any low-lying strength. Using the procedure described in section~\ref{sec:resp}, the main component of the trial input function consists of a Breit-Wigner distribution~\cite{die1988} used to describe the GDR. Since it is often more convenient to express the cross section in terms of E1 energy-weighted sum-rule strength, a fit parameter $S_{EWSR}$ is used to describe the fraction of the exhausted strength, in addition to the mean $E_{m}$ and width $\Gamma$ parameters. In addition to the GDR function, five individual square bins, with widths ranging from 0.5~MeV at the 1n threshold up to 2.8~MeV at an excitation energy of 13.7~MeV, were added to describe possible deviations from a Breit-Wigner shaped E1 distribution in the 1n channel. The bin width was adjusted to take into account the resolution of the excitation energy, based on the statistical decay simulations described in section~\ref{sec:resp}. Before starting the fit procedure, the previously described strength distribution is converted to the proper Coulomb excitation differential cross section using the virtual photon spectra, as illustrated by Fig.~\ref{fig:ni68_vps}.

\begin{figure}
 \centering
\begin{overpic}[scale=0.6]{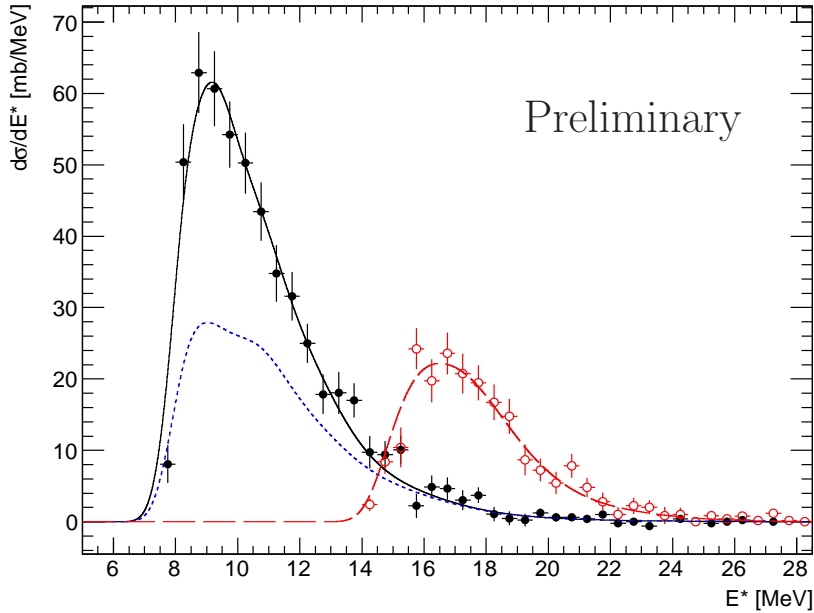}
\put(70,65){\begin{Large}Preliminary\end{Large}}
\end{overpic} 
\caption{(Color online) Preliminary Coulomb excitation cross section of $^{68}$Ni. The measured 1n (solid black points) and 2n (open red circles) data is shown, as well as the fit result for the 1n (solid black line) and 2n (dashed red line) channels. The dotted blue line shows the contribution of the GDR in the 1n data.}\label{fig:ni68_estar}
\end{figure}
The result of the fitting procedure is displayed in Fig.~\ref{fig:ni68_estar} for the measured Coulomb excitation cross section of $^{68}$Ni. The fit of the GDR distribution was performed on the 2n data, since the eventual low-lying E1 strength is not expected to influence this neutron channel in any way. The $\chi^{2}$ minimization leads to $E_{m}=18.1(5) $~MeV, $\Gamma=6.1(5) $~MeV and an E1 energy-weighted sum-rule strength $S_{EWSR}=97(9)$\%, results that are in good agreement with systematic values, such as those proposed by Junghans \textit{et al.}~\cite{jun2008}, predicting the GDR of $^{68}$Ni at 18.17~MeV with a width of 5.17~MeV. Due to the width of the GDR and its proximity to the 2n threshold, a substantial amount of cross section belonging to the GDR is expected to reside in the 1n channel. The strong distortion induced by the experimental response however shifts this strength to lower energies, as shown by the blue dotted line in Fig.~\ref{fig:ni68_estar}, also clearly showing an excess of E1 strength below the GDR. It should also be noted that the fit procedure is carried out simultaneously on several observables, in order to obtain a more global description of the experimental data than by fitting each observable individually. In the present case, the fit is performed on the excitation energy, the neutron kinetic energy and their longitudinal momentum, both for the 1n and 2n decay channels, and additionally on the sum of the kinetic energies of the neutrons for the 2n channel. For instance, the fits presented in Fig.~\ref{fig:ekin} and \ref{fig:ni68_estar} were obtained simultaneously.

A preliminary analysis of the additional E1 strength shows a Gaussian peak at 10.4(4)~MeV with a width of $\sigma=0.8(3)$~MeV. These parameters are in excellent agreement with the PDR measured by ($\gamma^{*}$,$\gamma'$) virtual photon scattering~\cite{wie2009}, reported at 11~MeV with a width of less than 1~MeV. In the present case, the peak integral corresponds to 4.1(1.9)\% of the E1 energy-weighted sum-rule strength. This is again in good agreement with the value of Wieland \textit{et al.} of 5\%~\cite{wie2009} obtained with a gamma decay branching ratio of 4\% for the entire PDR region.

\section{Summary and outlook}

The analysis of the E1 strength in $^{68}$Ni has been described, along with two important steps required to quantitatively extract its low-lying part. The first step consisted in evaluating the detector response of the R$^{3}$B-LAND setup, resulting in the calculation of response matrices, allowing comparisons between theoretical strength distributions with the measured data. The second step was the measurement of the branching ratio for the non-statistical decay of the compound nucleus, also improving the quality of the fit of convoluted input distributions to the data. Based on this information, a fit of a trial E1 strength function, consisting of a Breit-Wigner function and of five independent bins spanning the 1n decay channel, was carried out and revealed a GDR in good agreement with systematics, as well as a non-negligible amount of low-lying strength. This preliminary additional cross section showed a peak structure exhausting 4.1(1.9)\% of the E1 energy-weighted sum rule.

The observed PDR strength will provide a further constraint of the symmetry-energy parameters of the equation-of-state, similarly to Carbone \textit{et al.} for the virtual photon scattering data on $^{68}$Ni. This task will be performed using not only the E1 strength in the PDR region, but also using the dipole polarizability measured in the energy range starting at the neutron threshold and ending beyond the GDR. Since not only $^{68}$Ni was measured in this experiment, but also more exotic species such as $^{70}$Ni and $^{72}$Ni, a similar analysis will be applied to these isotopes, allowing a verification of the behavior of the symmetry energy at different asymmetry values.


\section*{References}
\begin{thebibliography}{99}
\bibitem{die1988} Dietrich S S and Berman B L, \textit{At. Data Nucl. Data Tables} \textbf{38}, 199 (1988).
\bibitem{pie2011} Piekarewicz J, \PR \textit{C} \textbf{83}, 034319 (2011).
\bibitem{gor1998} Goriely S, \PL \textit{B} \textbf{436}, 10 (1998).
\bibitem{adr2005} Adrich P \textit{et al.}, \PRL \textbf{95}, 132501 (2005).
\bibitem{kli2007} Klimkiewicz A \textit{et al.}, \PR \textit{C} \textbf{76}, 051603(R) (2007).
\bibitem{wie2009} Wieland O \textit{et al.}, \PRL \textbf{102}, 092502 (2009).
\bibitem{car2010} Carbone A \textit{et al.}, \PR \textit{C} \textbf{81}, 041301(R) (2010).
\bibitem{gei1992} Geissel H \textit{et al.}, \NIM \textit{B} \textbf{70}, 286 (1992).
\bibitem{cub1998} Cub J \textit{et al.}, \NIM \textit{A} \textbf{402}, 67 (1998).
\bibitem{bla1992} Blaich T \textit{et al.}, \NIM \textit{A} \textbf{314}, 136 (1992).
\bibitem{ber1986} Bertulani C A and Baur G, \NP \textit{A} \textbf{458}, 725 (1986).
\bibitem{ben1989} Benesh C J, Cook B C and Vary J P, \PR \textit{C} \textbf{40}, 1198 (1989).
\bibitem{ago2003} Agostinelli S \textit{et al.}, \NIM \textit{A} \textbf{506}, 250 (2003).
\bibitem{ber2008} Bertini D \textit{et al.}, \textit{J. Phys.: Conf. Series} \textbf{119}, 032011 (2008).
\bibitem{wei1937} Weisskopf V, \PR \textbf{52}, 295 (1937).
\bibitem{lec1959} Le Couteur K J, Lang D W, \NP \textbf{13}, 32 (1959).
\bibitem{kel2008} Kelic A, Ricciardi M and Schmidt K H, \textit{Proceedings of Joint ICTP-IAEA Advanced Workshop on Model Codes for Spallation Reactions, ICTP Trieste}, February 4-8 2008, arXiv:0906.4193v1.
\bibitem{gai1991} Gaimard J J and Schmidt K H, \NP \textit{A} \textbf{531}, 709 (1991).
\bibitem{har2001} Harakeh M N and van der Woude A, \textit{Giant Resonances; Fundamental High-Frequency Modes of Nuclear Excitation}, Clarendon Press, Oxford (2001).
\bibitem{jun2008} Junghans A R \textit{et al.}, \PL \textit{B} \textbf{670}, 200 (2008).
\end{thebibliography}
\end{document}